\documentclass[journal]{IEEEtran}
\usepackage[T1]{fontenc}
\usepackage{graphicx}
\usepackage{booktabs}
\usepackage{amsmath,amssymb}
\usepackage{array}
\usepackage{multirow}
\usepackage{url}
\usepackage{cite}
\usepackage{stfloats}
\usepackage{placeins}
\usepackage[svgnames]{xcolor}
\usepackage[hidelinks]{hyperref}
\graphicspath{{Figs_nplsd/}}
\usepackage{orcidlink}
\usepackage{subcaption}
\newcommand{\ignore}[1]{}
\newcommand{\sAP}[1]{\ensuremath{\mathrm{sAP}^{#1}}}

\begin{document}
	\title{NPLSD: Accelerating Line-Segment Detection on NPU Microcontrollers}
	\author{
		Parsa Hassani Shariat Panahi~\orcidlink{0009-0005-2912-3754},
		Amir Hossein Jalilvand~\orcidlink{0000-0002-7641-6606},
		and M. Hassan Najafi~\orcidlink{0000-0002-4655-6229} \\
		\thanks{Parsa Hassani Shariat Panahi and Amir Hossein Jalilvand are with the School of Computer Engineering, Iran University of Science and Technology, Tehran, Iran. M. Hassan Najafi is with the Electrical, Computer, and Systems Engineering Department, Case Western Reserve University, Cleveland, OH, USA. Corresponding Author:~M.~Hassan Najafi~najafi@case.edu}
	}
	
	\maketitle
	\begin{abstract}
		Line-segment detection is fundamental to robotics, autonomous navigation, and industrial inspection. While transformer-based detectors achieve the highest accuracy, their deployment on microcontrollers remains impractical due to resource constraints. The STM32N6, with its Neural-ART NPU, promises to enable deep vision at the extreme edge. However, existing detectors rely on attention, grid-sampling, and normalization, operators that are unsupported by the convolution-oriented NPU. This architectural mismatch is characterized operator by operator: attention, grid-sampling, and normalization lack accelerator primitives, and the decoder's self-attention alone materializes a 39 MB score tensor that exceeds on-chip memory. To address this limitation, NPLSD is introduced as a pair of NPU-compatible line-segment detectors built from one design methodology. NPLSD-H retains the convolutional HGNetv2 backbone of LINEA and replaces the transformer head with a fully-convolutional feature pyramid and an F-Clip dense head. NPLSD-M adapts the M-LSD-tiny trunk to the supported operator set. Warm-started from ImageNet and trained on ShanghaiTech Wireframe, the 2.63M-parameter NPLSD-H reaches $\mathrm{sAP}^{10}=37.9$ (35.9 int8); the 0.62M-parameter NPLSD-M reaches 41.9 (41.1 int8). A controlled ablation isolates the trunk as the only variable, and initialization alone accounts for 4.6 points.
	\end{abstract}
	
	\section{Introduction}\label{sec:intro}
	Line-segment detection (LSD) is a cornerstone of visual perception for \textit{robotics}, \textit{autonomous navigation}, and \textit{industrial metrology}.
	While classical methods such as the \textit{LSD} algorithm in~\cite{vongioi2010lsd} and the embedded-oriented \textit{ELSED} detector~\cite{suarez2021elsed} have long served as practical solutions, their accuracy is limited in challenging scenarios.
	LSD has been reshaped by deep learning in two major waves. The first wave introduced convolutional approaches including L-CNN~\cite{zhou2019lcnn}, HAWP~\cite{xue2020hawp,xue2023hawpv2}, and F-Clip~\cite{dai2021fclip}, which demonstrated the superiority of end-to-end learning over classical pipelines. DeepLSD~\cite{pautrat2023deeplsd} later combined learned gradient fields with classical refinement. The second wave brought transformer-based detectors such as LETR~\cite{xu2021letr}, DT-LSD~\cite{dtlsd}, and LINEA~\cite{linea}, which have established new state-of-the-art results. These models, however, are designed for GPU-class hardware. Even efficient convolutional variants like M-LSD~\cite{gu2022mlsd} and EM-LSD~\cite{em-lsd} target mobile SoCs with substantial memory resources.
	
	The \textit{embedded systems} landscape has shifted with the integration of on-chip neural processing units (NPUs) into microcontrollers. The STM32N6 exemplifies this trend. It pairs a Cortex-M55 with the Neural-ART NPU~\cite{stm32n6,neuralart}, a convolutional accelerator that raises throughput by more than an order of magnitude for supported operators. The ST Edge AI toolchain~\cite{stedgeai} provides a compilation pipeline for mapping convolutional neural networks onto this accelerator.
	
	Despite the promise of NPU-equipped microcontrollers, line-segment detection has remained absent from such platforms \cite{jalilvand2026lowlatency}. The primary impediment is \textit{architectural}. The most accurate detectors are transformer-based and rely on attention, grid-sampling, and normalization. These operators have no native support in a convolution-oriented NPU. They therefore execute on the host CPU. In the case of attention mechanisms, their memory footprint can exceed the on-chip activation budget entirely. Consequently, a transformer detector derives negligible benefit from the NPU.

	This mismatch is made concrete by analyzing LINEA-N against the accelerator's operator set. Only the \textit{convolutional backbone} is expressible in supported operators. The transformer encoder-decoder is not: the self-attention score tensor alone exceeds the available SRAM, and grid-sampling has no accelerator primitive. Whatever fraction of such a network a toolchain manages to compile, the transformer computation itself must execute on the host CPU, leaving the NPU idle.
	This gap motivates NPLSD, a pair of line-segment detectors designed for NPU-equipped microcontrollers. The design follows a single principle: \textit{every} operator must execute natively on the Neural-ART NPU. Two trunks are carried through the same methodology. NPLSD-H retains the convolutional HGNetv2-B0 backbone of LINEA and replaces the transformer head with a fully-convolutional feature pyramid and an F-Clip dense head; its graph contains only \texttt{Conv}, \texttt{ReLU}, \texttt{Add}, \texttt{Concat}, \texttt{Resize}, and \texttt{MaxPool}. NPLSD-M starts from the published M-LSD-tiny trunk, which cannot execute on the N6 as released, and adapts it: bilinear upsampling becomes nearest-neighbor resize, the TFLite-style asymmetric pads become symmetric convolution padding, the four-channel input becomes three, and ReLU6 is kept as a \texttt{Clip} node that the accelerator supports. All of these map onto the accelerator.
	
	Both detectors are warm-started from ImageNet and trained on \textit{ShanghaiTech} Wireframe under an identical head, loss, data, and schedule. NPLSD-H reaches $\mathrm{sAP}^{10}=37.9$; a from-scratch run of the same architecture reaches 33.3, so initialization alone accounts for 4.6 points. NPLSD-M reaches 41.9 with 4.2 times fewer parameters. Because only the trunk differs, the two models form a controlled ablation over trunk families. Since the accelerator is an integer engine, the configuration that matters is the quantized one. Both models are therefore exported to ONNX and quantized to int8 with static post-training quantization on real calibration images, and all results are reported for the fp32 and int8 graphs alike. Quantization costs 2.0 $\mathrm{sAP}^{10}$ points on the dense trunk and only 0.8 on the depthwise one.
	
	The contributions of this paper are as follows:
	\begin{itemize}
		\item The transformer-NPU gap is characterized at the operator level. Profiling LINEA-N against the Neural-ART instruction set reveals that only its backbone is expressible on the accelerator; the decoder cannot fit on-chip at all.
		
		\item Two NPU-native detectors, NPLSD-H and NPLSD-M, are introduced. Both networks are constructed entirely from the supported operator set and share an identical head and training recipe, enabling a controlled ablation of the trunk.
		
		\item ImageNet warm-starting improves $\mathrm{sAP}^{10}$ by 4.6 points at no architectural cost. The depthwise trunk loses only 0.8 points to int8 quantization, versus 2.0 for the dense trunk, while achieving higher accuracy with 4.2$\times$ fewer parameters.
	\end{itemize}
	The remainder of this paper is organized as follows. The transformer-NPU gap is characterized in Section~\ref{sec:gap}. The \textcolor{black}{NPLSD} architecture\textcolor{black}{s} and training are presented in Section~\ref{sec:method}, together with the int8 quantization procedure. Experimental results are reported in Section~\ref{sec:results}. Discussion and concluding remarks are provided in Sections~\ref{sec:discussion} and~\ref{sec:conclusion}.

	\section{\textcolor{black}{The NPU Gap: An Architectural Analysis}}\label{sec:gap}
	
	The characterization proceeds by establishing why transformer-based detectors cannot benefit from the STM32N6 NPU. Table~\ref{tab:ops} presents the operator mapping for LINEA-N, derived from its exported graph and the accelerator's documented operator support.
	
	The convolutional components of the HGNetv2-B0 backbone and the CNN cross-scale fusion in the hybrid encoder are mappable to the NPU. In contrast, the transformer-specific operators are not supported. GridSample, the bilinear sampling core of the line-attention deformable decoder, has no Neural-ART primitive and can only execute on the Cortex-M55. The same applies to LayerNorm, the attention Softmax, and the TopK and Gather operations for query selection.
	
	Critically, the decoder's self-attention operates over 1,100 queries and materializes an $[8,1100,1100]$ score tensor. This tensor is approximately 39 MB in floating-point, which exceeds the on-chip activation budget. The full decoder is therefore unplaceable. Only the backbone is expressible on the accelerator, and it represents a minority of the network's computation. The transformer-specific operations are consequently relegated to the host CPU. The accelerator is thus architecturally wasted on a transformer detector. This is the gap that NPLSD is designed to close.
	
	\begin{table}[]
		\caption{LINEA-N operator mapping on the STM32N6. NPU denotes the Neural-ART accelerator; CPU denotes the Cortex-M55 fallback.}
		\label{tab:ops}
		\centering
		\small
		\begin{tabular}{p{0.65\columnwidth} c}
			\toprule
			Block / operator & Target\\
			\midrule
			HGNetv2-B0 backbone (Conv, BN, ReLU, Pool) & NPU\\
			Encoder CNN fusion (Conv, Resize-nearest, Concat) & NPU\\
			Linear/Gemm projections (q/k/v, FFN) & NPU\\
			Attention Softmax & CPU\\
			LayerNorm & CPU\\
			GridSample (line-attention sampling) & CPU\\
			TopK + Gather (query selection) & CPU\\
			Decoder self-attention ($[8,1100,1100]$) & cannot place\\
			\bottomrule
		\end{tabular}
		\vspace{-1em}
	\end{table}
	\section{NPLSDH: Architecture and Training}\label{sec:method}
	
	This section presents the design and training of \textcolor{black}{the two NPLSD variants}. The architecture\textcolor{black}{s} comprise \textcolor{black}{ a trunk} for feature extraction, a lightweight \textcolor{black}{fusion stage}, and a \textcolor{black}{shared} dense head for line parameter regression. The training objective and optimization procedure are then formulated.
	
	\subsection{Design Principle}
	
	The design of NPLSDH is governed by a single, non-negotiable constraint: every operator must execute natively on the Neural-ART NPU. This requirement is not merely an optimization consideration; it is a fundamental precondition for the network to benefit from the STM32N6's accelerator at all. The operators supported by the NPU include convolution, batch normalization (folded into convolution during inference), ReLU, the bounded \texttt{Clip} activation (ReLU6), element-wise addition, concatenation, nearest-neighbor resize, and max-pool. Conversely, attention, grid-sampling, layer normalization, softmax, and top-k selection are explicitly excluded from the NPU's instruction set. Both NPLSD variants are therefore constructed exclusively from the supported operator set.
	
	\subsection{\textcolor{black}{NPLSD-H} Architecture}
	The NPLSDH architecture comprises three main components: backbone, neck, and head.
	
	\begin{itemize}
		\item \textit{\textbf{Backbone}}:  HGNetv2-B0~\cite{peng2024dfine,zhao2024rtdetr}, the convolutional backbone used by LINEA, is adopted. It produces feature maps at strides 8, 16, and 32 with 256, 512, and 1024 channels, respectively. Unlike the detection-transformer usage, ordinary batch normalization is kept rather than frozen affine layers. This ensures that during quantization, each batch normalization layer folds cleanly into its preceding convolution. The backbone is warm-started from the publicly available ImageNet-pretrained HGNetv2-B0 weights, with the optional affine layers dropped; every remaining backbone tensor is initialized from the checkpoint. A from-scratch run of the identical architecture is kept as an ablation.
		
		\item \textit{\textbf{Neck}}: A lightweight convolutional feature pyramid fuses the three backbone scales top-down. It uses 1x1 lateral convolutions, nearest-neighbor 2x upsampling (a supported NPU resize operation), and 3x3 smoothing convolutions. An additional upsampling stage produces a single stride-4 map. For a 320x320 input, this yields an 80x80 grid for the dense head.
		
		\item \textit{\textbf{Head}}: An F-Clip-style dense line head~\cite{dai2021fclip} predicts a six-channel output per grid cell. This includes a line-center likelihood (one channel), a sub-pixel center offset (two channels), a half-length (one channel), and a double-angle orientation (two channels). The double-angle encoding, defined as $(\cos 2\theta, \sin 2\theta)$, removes the 180-degree direction ambiguity inherent in line segments. The head emits raw feature maps. The center map is decoded off the NPU through a lightweight arithmetic pass comprising sigmoid activation, 3x3 max-pool non-maximum suppression, and per-peak endpoint reconstruction.
	\end{itemize}
	
	The complete network has 2.63M parameters, smaller than LINEA-N's 3.9M. Its exported graph contains only \texttt{Conv}, \texttt{ReLU}, \texttt{Add}, \texttt{Concat}, \texttt{Resize}, and \texttt{MaxPool} nodes, all of which map onto the accelerator.
	
	\subsection{NPLSD-M: Adapting the M-LSD Trunk}
	
	On mobile SoCs, M-LSD-tiny remains the reference lightweight detector, reporting $\sAP{10}=58.0$ at 512-pixel input with 0.6M parameters~\cite{gu2022mlsd}. Direct reuse of this trunk on the N6 is not feasible. The released model is float-only TFLite, whereas the integer accelerator does not execute floating-point convolutions. Its bilinear upsampling with aligned corners lacks a Neural-ART primitive, and the TFLite-style asymmetric padding triggers a code-generation fault. Furthermore, at 512-pixel resolution, a single early tensor of $96\times256\times256$ occupies 6.3 MB in int8, exceeding the N6's 4.2 MB internal activation SRAM.
	
	NPLSD-M is the accelerator-native redesign of this trunk, produced by the same methodology as NPLSD-H. The MobileNetV2 inverted-residual stack (blocks 0 through 10, taps at strides 4, 8, and 16) is warm-started from ImageNet. The A/B fusion blocks of M-LSD are retained, with their bilinear interpolation replaced by nearest-neighbor resize. Asymmetric pads become symmetric convolution padding, the four-channel input becomes standard RGB, and the M-LSD prediction head is replaced by the same F-Clip dense head used in NPLSD-H, at 320-pixel input. ReLU6 activations are retained as \texttt{Clip} nodes, which belong to the accelerator's supported operator set. The exported graph holds 44 \texttt{Conv}, 20 \texttt{Clip}, 10 \texttt{ReLU}, 8 \texttt{Add}, 3 \texttt{Concat}, and 2 \texttt{Resize} nodes, with no \texttt{Pad}, for 0.62M parameters and 1.74 GMACC.
	
	Since trunk choice is the only difference between the two variants, their comparison constitutes a controlled ablation of trunk families under a fixed head, loss, dataset, resolution, and schedule.
	
	\subsection{Training}
	
	Both variants are trained on the ShanghaiTech Wireframe benchmark. Ground-truth segments are rasterized into dense targets. A CenterNet-style Gaussian is placed on the center map at each ground-truth line center. Regression targets for offset, length, and angle are assigned to each positive center cell.
	
	The loss function combines a penalty-reduced focal loss on the center map with L1 or smooth-L1 regression losses on the offset, length, and unit-normalized angle at positive cells. The total loss is defined as:
	
	\begin{equation}
		\mathcal{L} = \mathcal{L}_{center} + \lambda_{reg} \left( \mathcal{L}_{offset} + \mathcal{L}_{length} + \mathcal{L}_{angle} \right)
	\end{equation}
	where $\lambda_{reg}$ balances the classification and regression terms.
	
	Optimization is performed with AdamW using a cosine annealing learning rate schedule. Each model is trained for 120 epochs with a batch size of 16, selecting the checkpoint with the best validation $\sAP{10}$. Input images are resized to 320x320. Evaluation is conducted using structural average precision ($\sAP{}$) on the 128x128 reference space following the standard Wireframe benchmark protocol.
	
	Warm-started from ImageNet, NPLSD-H achieves $\mathrm{sAP}^{10}=37.9$, while NPLSD-M reaches 41.9. Training NPLSD-H from scratch under the same recipe yields 33.3, indicating that ImageNet initialization contributes 4.6 points without any architectural modification. Notably, the accelerator operator set does not constitute the accuracy bottleneck; holding the graph fixed and varying only the initialization produces a larger accuracy shift than the gap between the two trunk families. Both results remain below transformer-based detectors, which is consistent with compact convolutional models evaluated at 320-pixel input.
	
	\begin{figure}[]
		\centering
		\includegraphics[width=1\columnwidth, trim=0.5cm 0cm 0.5cm 0cm, clip]{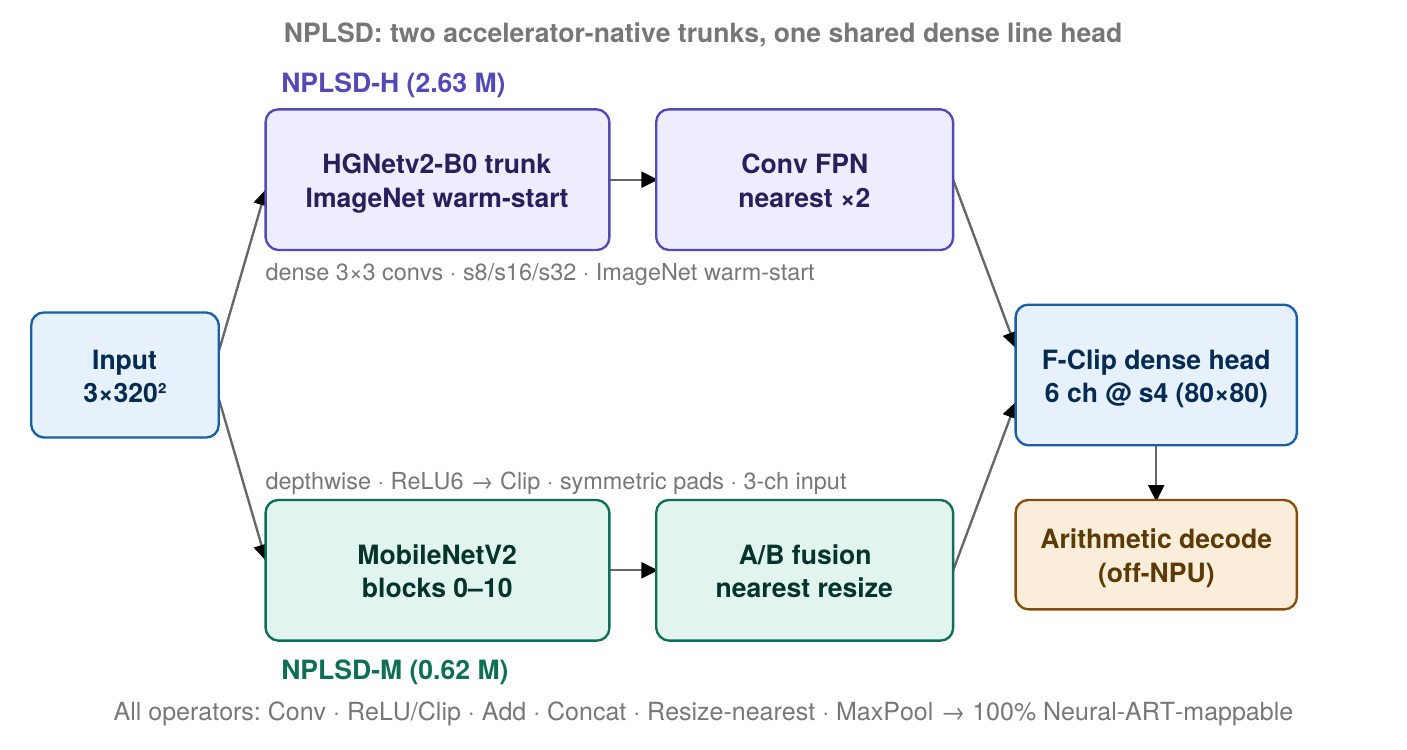}
		\caption{NPLSD architecture comprises two accelerator-native trunks and a shared dense head. The dense HGNetv2-B0 of NPLSD-H and the depthwise MobileNetV2 stack of NPLSD-M feed a six-channel F-Clip head at stride 4, followed by an arithmetic decode. Every operator (\texttt{Conv}, \texttt{ReLU/Clip}, \texttt{Add}, \texttt{Concat}, \texttt{Resize}, \texttt{MaxPool}) maps onto the Neural-ART NPU.}
		\label{fig:arch}
		\vspace{-2em}
	\end{figure}
	
	\subsection{Export and int8 Quantization}\label{sec:deploy}
	Since the Neural-ART is an integer engine, the configuration of record is the quantized one, and it is prepared entirely off-device. Each trained model is exported to ONNX at 320x320 input resolution and simplified to static shapes, yielding a clean graph over the supported operator set (the NPLSD-M graph is exported with symmetric padding from the start and contains no \texttt{Pad} node). Static int8 quantization~\cite{jacob2018quant} is then applied in QDQ (Quantize-Dequantize) format with per-channel weights, calibrated on 200 real Wireframe images; for NPLSD-M the bounded \texttt{Clip} activations are included in the quantized set. Quantization compresses the weights roughly fourfold, to 2.55\,MB for NPLSD-H and 0.65\,MB for NPLSD-M. All int8 results in Section~\ref{sec:results} are produced by executing these quantized graphs in ONNX Runtime over the full validation split. The training notebooks and the evaluation code reproducing every accuracy figure are publicly available on Kaggle~\cite{linpu_kaggle,linpu_warm_kaggle,mlsdn6_kaggle}, together with the dataset and the ImageNet-pretrained trunk checkpoints~\cite{lsd_dst_kaggle}.
	
	\section{Results}\label{sec:results}
	\subsection{Qualitative Detections}
\begin{figure}[]
		\centering
		\includegraphics[width=\columnwidth, trim=0cm 2.5cm 0cm 3cm, clip]{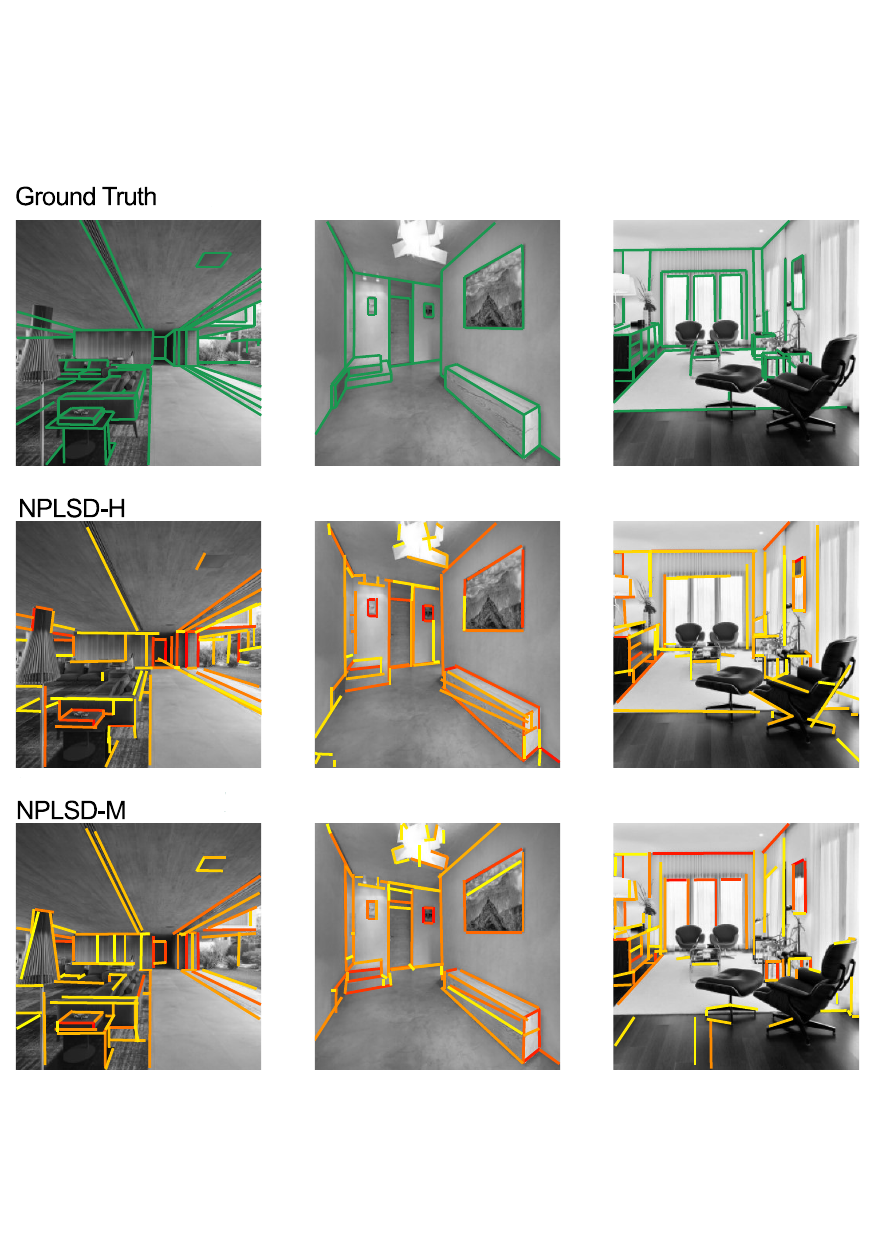}
		\caption{Qualitative NPLSD detections on ShanghaiTech Wireframe. Top: ground-truth. Middle: NPLSD-H. Bottom: NPLSD-M, predictions colored by confidence. Both models capture dominant structural lines while missing finer detail, with NPLSD-M recovering more of the long edges.}
		\label{fig:qual}
		\vspace{-1em}
	\end{figure}

	Fig. ~\ref{fig:qual} shows \textcolor{black}{NPLSD} 
	predictions on held-out ShanghaiTech Wireframe validation images spanning diverse indoor and outdoor man-made scenes. Across bedrooms, living rooms, facades, and terraces, the model\textcolor{black}{s} recover the dominant scene structure, including wall, ceiling, window, door, and furniture edges together with the perspective convergence of facade and floor boundaries. These predictions are generated directly by the fully convolutional, accelerator-mappable network\textcolor{black}{s}. Fine, low-contrast, and heavily occluded segments are more frequently missed, with NPLSD-M recovering visibly more of the long structural edges, in line with its higher per-length recall (Table~\ref{tab:length}). The results show the complete set of raw decoded outputs rather than a curated subset, providing qualitative evidence that native line-segment detection on a microcontroller NPU can produce coherent wireframe reconstructions.

	\subsection{Accuracy on the Wireframe Benchmark}
\begin{figure}[]
		\centering
		\includegraphics[width=0.9\columnwidth]{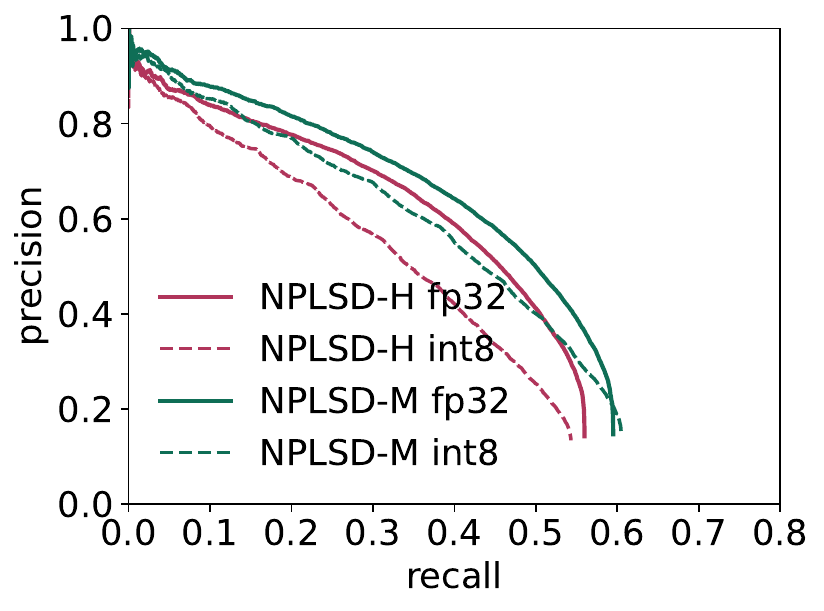}
		\caption{Precision-recall curves at the $\mathrm{sAP}^{10}$ matching threshold over the full validation split: NPLSD-H fp32/int8 (solid/dashed, $\mathrm{sAP}^{10}=37.9/35.9$) and NPLSD-M fp32/int8 ($41.9/41.1$). Real-calibration quantization shifts each curve only marginally; the int8 models are, for accuracy purposes, the same as their fp32 references.}
		\label{fig:pr}
		\vspace{-1em}
	\end{figure}

	Table~\ref{tab:acc} lists structural average precision over the complete 462-image validation split, evaluated at the three standard matching thresholds. Six configurations are compared: each variant in fp32 and int8, together with the from-scratch NPLSD-H ablation pair. Warm-started NPLSD-H reaches $\mathrm{sAP}^{10}=37.9$; after static quantization with 200 real calibration images (QDQ format, per-channel weights), the int8 network intended for the accelerator retains 35.9. NPLSD-M reaches 41.9 and maintains 41.1 in int8, a loss of 0.8 points compared with 2.0 for the dense trunk. Degradations of this order are typical when compact convolutional networks are quantized to int8~\cite{obszarski2025quant}. Precision and recall are affected almost uniformly, as the curves in Fig.~\ref{fig:pr} indicate, so accuracy statements made for the fp32 model carry over to the quantized one. At a confidence threshold of 0.15, the warm NPLSD-H proposes 129 segments per image on average against 74.2 annotated (NPLSD-M: 165.5), a proposal surplus that MiLSD~\cite{milsd2026} also exhibits and that the ranked sAP protocol absorbs.

	\begin{table}[]
		\caption{Structural AP on the full Wireframe validation split (462 images). The int8 models are quantized with 200 real calibration images and are the configurations intended for the Neural-ART NPU (executed in ONNX Runtime). The from-scratch rows hold the NPLSD-H architecture fixed and vary only the initialization.}
		\label{tab:acc}
		\centering
		\small
		\setlength{\tabcolsep}{7pt}
		\begin{tabular}{lccc}
			\toprule
			Configuration & $\mathrm{sAP}^{5}$ & $\mathrm{sAP}^{10}$ & $\mathrm{sAP}^{15}$\\
			\midrule
			NPLSD-H fp32, from scratch   & 24.1 & 33.3 & 38.2\\
			NPLSD-H int8, from scratch   & 23.2 & 32.0 & 36.3\\
			NPLSD-H fp32, warm           & 28.5 & 37.9 & 42.3\\
			NPLSD-H int8, warm           & 26.8 & 35.9 & 40.3\\
			NPLSD-M fp32, warm           & 32.7 & 41.9 & 46.1\\
			NPLSD-M int8, warm           & 32.2 & 41.1 & 45.1\\
			\bottomrule
		\end{tabular}
	\end{table}

	For context, Table~\ref{tab:platform} situates these results among published line detectors across hardware classes. GPU-class detectors such as HAWP~\cite{xue2020hawp} at 66.5 and L-CNN~\cite{zhou2019lcnn} at 62.8 assume workstation memory budgets. M-LSD-tiny~\cite{gu2022mlsd} brought learned detection to mobile SoCs at 58.0, yet still relies on gigabytes of RAM. Below the mobile tier, the literature is, to our knowledge, empty. The only microcontroller results available are the 25k F-Clip baseline (10.6), MiLSD (24.1)~\cite{milsd2026}, and the present work, all three originating from the same research effort, reflecting the novelty of this design space. Within this comparison, NPLSD-H improves $\mathrm{sAP}^{10}$ by 11.8 points over MiLSD, and NPLSD-M by 17.0 points, with both networks constructed entirely from the accelerator's supported operator set. NPLSD-M also closes 41 percent of the gap separating MiLSD from the mobile-SoC M-LSD-tiny, while remaining int8 and microcontroller-scale at 320-pixel input. For reference, the classical LSD algorithm of von Gioi et al.~\cite{vongioi2010lsd}, evaluated on a desktop CPU, attains an $\mathrm{sAP}^{10}$ of 8.8. The int8 NPLSD-M model achieves approximately 4.7 times this score while targeting substantially more resource-constrained hardware.

	\begin{table}[]
		\caption{Wireframe $\mathrm{sAP}^{10}$ comparison across hardware classes. GPU and mobile results are from the M-LSD study~\cite{gu2022mlsd} and the original papers (512-pixel inputs). Microcontroller entries are from this work and MiLSD~\cite{milsd2026}.}
		\label{tab:platform}
		\centering
		\small
		\setlength{\tabcolsep}{5pt}
		\begin{tabular}{lccl}
			\toprule
			Method & $\mathrm{sAP}^{10}$ & Params & Platform\\
			\midrule
			LSD~\cite{vongioi2010lsd}            & 8.8  & ---     & desktop CPU\\
			TP-LSD-Lite~\cite{huang2020tplsd}    & 59.7 & 23.9M & GPU\\
			L-CNN~\cite{zhou2019lcnn}            & 62.8 & 9.8M  & GPU\\
			LINEA-N~\cite{linea}                 & 65.0 & 3.9M  & GPU\\
			HAWP~\cite{xue2020hawp}              & 66.5 & 10.4M & GPU\\
			M-LSD-tiny~\cite{gu2022mlsd}         & 58.0 & 0.6M  & mobile SoC\\
			\midrule
			F-Clip-25k                           & 10.6 & 0.025M & MCU (M7)\\
			MiLSD~\cite{milsd2026}               & 24.1 & 0.39M  & MCU (M7)\\
			NPLSD-H int8 (ours)          & 35.9 & 2.63M & MCU (M55+NPU)\\
			NPLSD-M int8 (ours) & 41.1 & 0.62M & MCU (M55+NPU)\\
			\bottomrule
		\end{tabular}
		\vspace{-1em}
	\end{table}

	\subsection{Cross-Dataset Generalization: YorkUrban}
	
	Following the standard protocol established by L-CNN, HAWP, F-Clip, LETR, and M-LSD~\cite{zhou2019lcnn,xue2020hawp,dai2021fclip,xu2021letr,gu2022mlsd}, which train on Wireframe and report YorkUrban as a zero-shot test, both NPLSD variants are evaluated on the 102 YorkUrban images without any fine-tuning (Table~\ref{tab:york}). All published detectors exhibit a sharp drop on this transfer; GPU-class models retain 42 to 45 percent of their Wireframe $\mathrm{sAP}^{10}$, while M-LSD-tiny retains 42 percent. NPLSD-M retains 41 percent (17.2 of 41.9), indicating that the accelerator-constrained design does not compromise cross-dataset generalization. The observed drop is expected: YorkUrban contains outdoor scenes and annotates only lines along the dominant Manhattan directions, with 118.8 ground-truth lines per image. Out of domain, quantization costs NPLSD-M only 0.3 points.
	
	\begin{table}[]
		\caption{Zero-shot structural AP on YorkUrban (102 images; trained on Wireframe only, no fine-tuning).}
		\label{tab:york}
		\centering
		\small
		\setlength{\tabcolsep}{7pt}
		\begin{tabular}{lccc}
			\toprule
			Configuration & $\mathrm{sAP}^{5}$ & $\mathrm{sAP}^{10}$ & $\mathrm{sAP}^{15}$\\
			\midrule
			NPLSD-H fp32 & 9.6 & 13.7 & 16.0\\
			NPLSD-H int8 & 8.5 & 12.5 & 14.5\\
			NPLSD-M fp32 & 13.0 & 17.2 & 19.6\\
			NPLSD-M int8 & 13.0 & 16.9 & 19.2\\
			\bottomrule
		\end{tabular}
		\vspace{-1em}
	\end{table}
	
	Recall stratified by ground-truth length appears in Table~\ref{tab:length}. For the warm NPLSD-H, short and medium segments are recovered at 54.0 and 62.2 percent, and segments longer than 25 units of the $128^2$ label space at 44.0 percent. NPLSD-M leads across all buckets at 61.0, 65.5, and 48.0 percent, respectively. Compared with the from-scratch model, pretraining improves long-segment recall by 5.9 points, the largest per-bucket gain. A long-line deficit nevertheless persists. This pattern is consistent with expectations. MiLSD, built on the same center, length, and angle encoding, exhibits similar behavior~\cite{milsd2026}; thus, the effect appears to arise from the representation family rather than from the NPU constraint introduced here. Multi-scale heads would be a natural extension and remain compatible with the accelerator's operator set.

	\begin{table}[]
		\caption{Recall by ground-truth segment length (fp32, confidence above 0.15, $\mathrm{sAP}^{10}$ matching distance; lengths in $128\times128$ label space). Warm-starting improves long-segment recall most; NPLSD-M leads across all buckets. The reduced recall on long segments is consistent with MiLSD~\cite{milsd2026}, which uses the same center-based encoding.}
		\label{tab:length}
		\centering
		\small
		\setlength{\tabcolsep}{5pt}
		\begin{tabular}{lcccc}
			\toprule
			GT length & \#\,GT & H scratch & H warm & M warm\\
			\midrule
			short ($<10$)        & 11,997 & 55.1\% & 54.0\% & 61.0\%\\
			medium ($10$--$25$)  & 12,195 & 58.5\% & 62.2\% & 65.5\%\\
			long ($>25$)         & 10,094 & 38.1\% & 44.0\% & 48.0\%\\
			\bottomrule
		\end{tabular}
		\vspace{-1em}
	\end{table}

	\begin{figure*}[]
		\centering
		\includegraphics[width=0.9\textwidth]{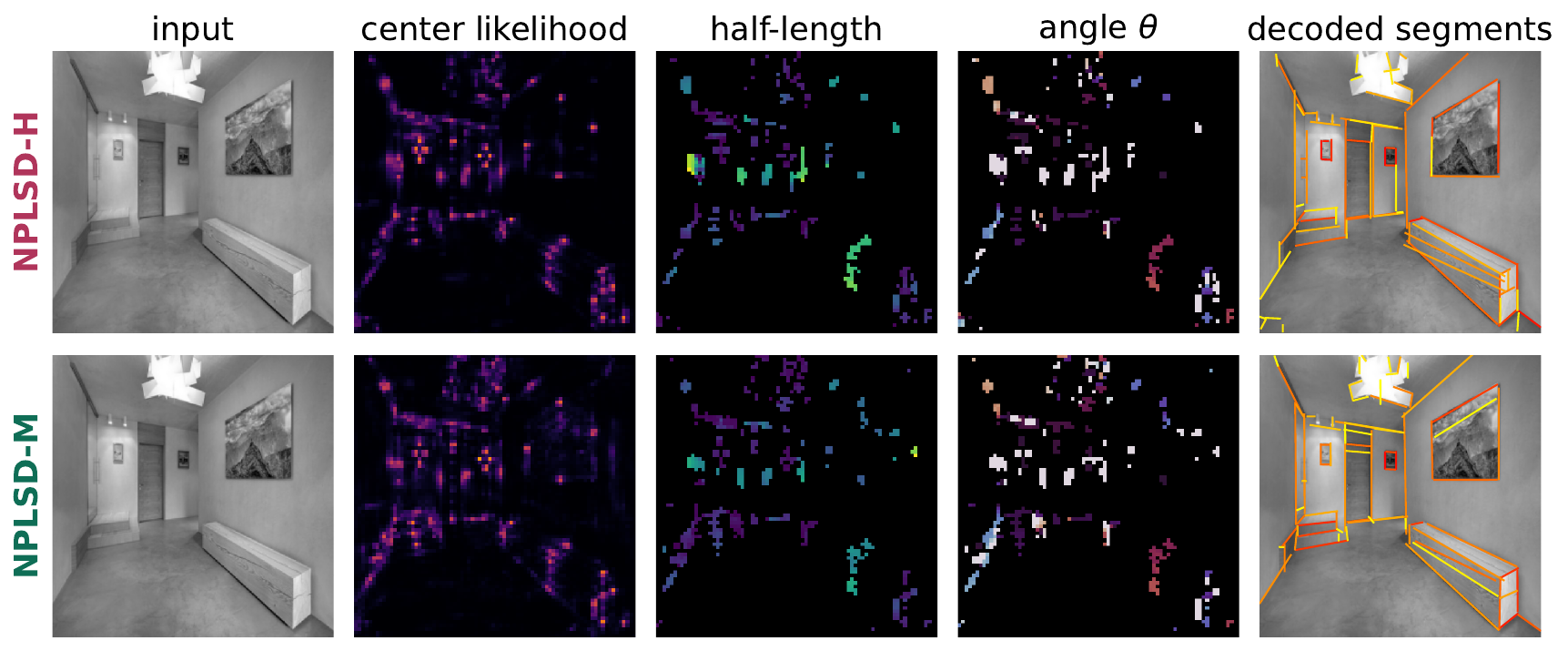}
		\caption{Output of the six-channel dense head for validation image \#170 (NPLSD-H top, NPLSD-M bottom). From left: input; raw center likelihood; half-length and double-angle orientation $\theta=\frac{1}{2}\,\mathrm{atan2}(\sin 2\theta,\cos 2\theta)$, both masked to the cells the decoder reads (center confidence above 0.15); segments reconstructed by the arithmetic decode, colored by confidence.}
		\label{fig:maps}
	\end{figure*}
	
	The behavior of the dense head can be characterized through direct examination of its output maps. Fig. ~\ref{fig:maps} renders the raw $80\times80$ output channels for one validation scene: center likelihood peaks at segment midpoints, the half-length map grows along extended structural edges, and the decoded orientation stays piecewise constant across surfaces that share a direction. Precisely these three quantities are assembled into segments by the \textcolor{black}{arithmetic} decode.
	
	\subsection{Model Footprint}
	
\begin{figure}[]
		\centering
		\includegraphics[width=\columnwidth, trim=0cm 0 0 1cm, clip]{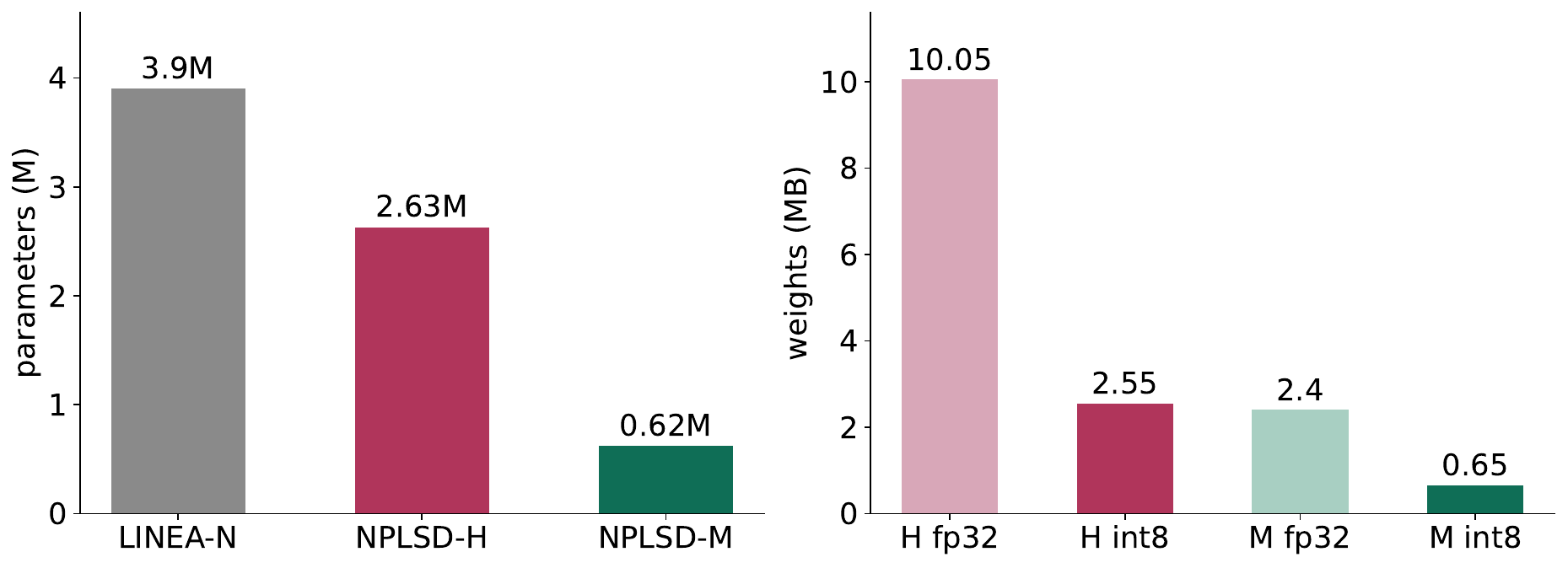}
		\caption{Parameter count and weight footprint comparison. Left: NPLSD-H has 2.63M parameters, smaller than LINEA-N's 3.9M, and NPLSD-M only 0.62M. Right: int8 quantization reduces NPLSD-H's weight footprint from 10.05 MB to 2.55 MB (NPLSD-M: 0.65 MB), a microcontroller-class storage budget.}
		\label{fig:size}
		\vspace{-1em}
	\end{figure}

	NPLSD-H has 2.63M parameters and 2.69 GMACC at 320x320; NPLSD-M has 0.62M parameters and 1.74 GMACC. Quantized to int8, the NPLSD-H weights occupy 2.55 MB, compared with 10.05 MB in floating-point; NPLSD-M compresses to 0.65 MB (Fig.~\ref{fig:size}). Both quantized weight files fit comfortably within microcontroller-class flash budgets, roughly four times smaller than their float counterparts. Notably, the 0.65 MB NPLSD-M is smaller than MiLSD-class detectors while achieving 17 points higher in $\mathrm{sAP}^{10}$.
	
	\section{Discussion and Limitations}\label{sec:discussion}
	
	NPLSD demonstrates that line-segment detection can be made compatible with a microcontroller NPU through co-design with the accelerator's operator set. The controlled trunk ablation carries its own lesson. With head, loss, data, and schedule held fixed, the depthwise MobileNetV2 trunk outscores the dense HGNetv2 trunk by 5.2 int8 points at 4.2 times fewer parameters, and it is markedly more robust to quantization. Both variants are nevertheless retained. Depthwise and dense convolutions place very different demands on an accelerator's MAC array and memory system; accuracy per parameter need not translate into accuracy per millisecond. Which trunk wins on silicon is a question this paper deliberately leaves to hardware evaluation.
	
	Two limitations frame the contribution. First, accuracy: at $\mathrm{sAP}^{10}=41.1$ in int8, NPLSD-M still trails transformer-based detectors and the 512-pixel mobile M-LSD-tiny. Knowledge distillation~\cite{mirzadeh2020takd} and multi-scale heads are natural next steps, and warm-starting, previously listed as future work, is now measured at +4.6 points. Second, evaluation covers two datasets at a fixed resolution, and all results are simulation-level (ONNX Runtime). On-silicon latency, memory placement, and energy are outside the scope of this paper. Despite these limitations, NPLSD provides a baseline and methodology for NPU-ready geometric vision on microcontrollers.
	
	\section{Conclusion}\label{sec:conclusion}
	
	This paper presented NPLSD, a pair of line-segment detectors designed for an NPU-equipped microcontroller. Motivated by an architectural gap, in which transformer detectors are largely inexpressible in a convolution accelerator's operator set, both variants replace non-convolutional structure with a fully-convolutional design. By construction, every operator belongs to the Neural-ART supported set.
	
	Warm-started on ShanghaiTech Wireframe, the 2.63M-parameter NPLSD-H reaches $\mathrm{sAP}^{10}=37.9$ (35.9 int8) and the 0.62M-parameter NPLSD-M reaches 41.9 (41.1 int8), 17.0 points above the best prior microcontroller result. Static int8 quantization, the format the integer accelerator requires, costs under one point on the depthwise trunk. The controlled ablation shows that initialization (+4.6) matters as much as trunk choice (+5.2), and that neither is limited by the accelerator operator set. NPLSD establishes that accelerator-compatible line-segment detection on microcontrollers is feasible at useful accuracy. Closing the remaining accuracy gap through knowledge distillation, and characterizing both variants on silicon, remain future work.
	\bibliographystyle{IEEEtran}
	\bibliography{References_nplsd}

@article{dai2021fclip,
  author  = {Dai, Xili and Gong, Haigang and Wu, Shuai and Yuan, Xiaojun and Ma, Yi},
  title   = {Fully Convolutional Line Parsing},
  journal = {Neurocomputing}, volume = {506}, pages = {1--11}, year = {2022},
  doi     = {10.1016/j.neucom.2022.07.026}, note = {F-Clip. arXiv:2104.11207}
}

@InProceedings{dtlsd,
    author    = {Janampa, Sebastian and Pattichis, Marios},
    title     = {DT-LSD: Deformable Transformer-Based Line Segment Detection},
    booktitle = {Proceedings of the Winter Conference on Applications of Computer Vision (WACV)},
    month     = {February},
    year      = {2025},
    pages     = {3477-3486}
}

@article{em-lsd,
title = {EM-LSD: A lightweight and efficient model for multi-scale line segment detection},
journal = {Robotics and Autonomous Systems},
volume = {195},
pages = {105192},
year = {2026},
issn = {0921-8890},
doi = {10.1016/j.robot.2025.105192},
url = {https://www.sciencedirect.com/science/article/pii/S0921889025002891},
author = {Shuo Hu and Liye Zhao and Qing Wang},
}

@inproceedings{gu2022mlsd,
  author    = {Gu, Geonmo and Ko, Byungsoo and Go, SeoungHyun and Lee, Sung-Hyun and Lee, Jingeun and Shin, Minchul},
  title     = {Towards Light-weight and Real-time Line Segment Detection},
  booktitle = {AAAI Conf. on Artificial Intelligence}, year = {2022},
  note = {M-LSD / M-LSD-tiny; MobileNetV2, center+displacement. arXiv:2106.00186}
}

@misc{linea,
      title={LINEA: Fast and Accurate Line Detection Using Scalable Transformers}, 
      author={Sebastian Janampa and Marios Pattichis},
      year={2025},
      eprint={2505.16264},
      archivePrefix={arXiv},
      primaryClass={cs.CV},
      url={https://arxiv.org/abs/2505.16264}, 
}

@misc{neuralart,
  author       = {{STMicroelectronics}},
  title        = {{ST Neural-ART Accelerator}: Introduction},
  howpublished = {\url{https://www.st.com/resource/en/product_presentation/st-neural-art-accelerator-introduction.pdf}},
  year         = {2025},
  note         = {Convolution-oriented NPU; epoch-controller execution model.}
}

@misc{peng2024dfine,
  author       = {Peng, Yansong and others},
  title        = {{D-FINE}: Redefine Regression Task in {DETRs} as Fine-grained Distribution Refinement},
  year         = {2024},
  howpublished = {arXiv:2410.13842},
  url          = {https://arxiv.org/abs/2410.13842},
  note         = {Source of the HGNetv2-B0 backbone used by LINEA and adapted here.}
}

@misc{stedgeai,
  author       = {{STMicroelectronics}},
  title        = {{ST Edge AI Suite / X-CUBE-AI / ST Edge AI Developer Cloud}},
  howpublished = {\url{https://www.st.com/en/embedded-software/x-cube-ai.html}},
  year         = {2025},
  note         = {Model analysis, int8 code generation, and on-target benchmarking for STM32.}
}

@misc{stm32n6,
  author       = {{STMicroelectronics}},
  title        = {{STM32N6} Series: Arm Cortex-M55 Microcontrollers with Neural-ART Accelerator},
  howpublished = {\url{https://www.st.com/en/microcontrollers-microprocessors/stm32n6-series.html}},
  year         = {2025},
  note         = {Neural-ART neural processing unit; on-chip SRAM $\sim$4.2\,MB; external xSPI flash and hyperRAM.}
}

@article{vongioi2010lsd,
  author  = {von Gioi, Rafael Grompone and Jakubowicz, J{\'e}r{\'e}mie and Morel, Jean-Michel and Randall, Gregory},
  title   = {{LSD}: A Fast Line Segment Detector with a False Detection Control},
  journal = {IEEE Trans. on Pattern Analysis and Machine Intelligence (TPAMI)},
  volume  = {32}, number = {4}, pages = {722--732}, year = {2010}
}

@inproceedings{xu2021letr,
  author    = {Xu, Yifan and Xu, Weijian and Cheung, David and Tu, Zhuowen},
  title     = {Line Segment Detection Using Transformers without Edges},
  booktitle = {IEEE/CVF Conf. on Computer Vision and Pattern Recognition (CVPR)},
  year      = {2021}, note = {LETR. arXiv:2101.01909}
}

@inproceedings{xue2020hawp,
  author    = {Xue, Nan and Wu, Tianfu and Bai, Song and Wang, Fu-Dong and Xia, Gui-Song and Zhang, Liangpei and Torr, Philip H.S.},
  title     = {Holistically-Attracted Wireframe Parsing},
  booktitle = {IEEE/CVF Conf. on Computer Vision and Pattern Recognition (CVPR)},
  year      = {2020}, note = {HAWP. arXiv:2003.01663}
}

@article{xue2023hawpv2,
  author  = {Xue, Nan and Wu, Tianfu and Bai, Song and Wang, Fu-Dong and Xia, Gui-Song and Zhang, Liangpei and Torr, Philip H.S.},
  title   = {Holistically-Attracted Wireframe Parsing: From Supervised to Self-Supervised Learning},
  journal = {IEEE Trans. on Pattern Analysis and Machine Intelligence (TPAMI)},
  year    = {2023}, note = {HAWPv2/v3. arXiv:2210.12971}
}

@inproceedings{zhao2024rtdetr,
  author    = {Zhao, Yian and Lv, Wenyu and Xu, Shangliang and Wei, Jinman and Wang, Guanzhong and Dang, Qingqing and Liu, Yi and Chen, Jie},
  title     = {{DETRs} Beat {YOLOs} on Real-time Object Detection},
  booktitle = {IEEE/CVF Conf. on Computer Vision and Pattern Recognition (CVPR)},
  year      = {2024}, note = {RT-DETR. arXiv:2304.08069}
}

@inproceedings{zhou2019lcnn,
  author    = {Zhou, Yichao and Qi, Haozhi and Ma, Yi},
  title     = {End-to-End Wireframe Parsing},
  booktitle = {IEEE/CVF Int. Conf. on Computer Vision (ICCV)},
  year      = {2019}, note = {L-CNN. arXiv:1905.03246}
}

@misc{milsd2026,
  author        = {Hassani Shariat Panahi, Parsa and Jalilvand, Amir H. and Najafi, M. Hassan},
  title         = {{MiLSD}: Micro Line-Segment Detector},
  year          = {2026},
  howpublished  = {arXiv:2607.06600},
  url           = {https://arxiv.org/abs/2607.06600},
  note          = {Companion work; same center/length/angle representation on a no-NPU Cortex-M7.}
}

@inproceedings{huang2020tplsd,
  author    = {Huang, Siyu and Qin, Fangbo and Xiong, Pengfei and Ding, Ning and He, Yijia and Liu, Xiao},
  title     = {{TP-LSD}: Tri-Points Based Line Segment Detector},
  booktitle = {European Conf. on Computer Vision (ECCV)}, year = {2020}, note = {arXiv:2009.05505}
}

@misc{linpu_kaggle,
  author       = {Hassani Shariat Panahi, Parsa},
  title        = {{LI-NPU}: Training and Deployment Notebook},
  year         = {2026},
  howpublished = {\url{https://www.kaggle.com/code/parsahshariatpanahi/linpu}},
  note         = {Kaggle notebook: model, training, structural-AP evaluation, and ONNX export.}
}

@inproceedings{jacob2018quant,
  author    = {Jacob, Benoit and Kligys, Skirmantas and Chen, Bo and Zhu, Menglong and Tang, Matthew and Howard, Andrew and Adam, Hartwig and Kalenichenko, Dmitry},
  title     = {Quantization and Training of Neural Networks for Efficient Integer-Arithmetic-Only Inference},
  booktitle = {IEEE/CVF Conf. on Computer Vision and Pattern Recognition (CVPR)}, year = {2018}, note = {arXiv:1712.05877}
}

@inproceedings{pautrat2023deeplsd,
  author    = {Pautrat, R{\'e}mi and Barath, Daniel and Larsson, Viktor and Oswald, Martin R. and Pollefeys, Marc},
  title     = {{DeepLSD}: Line Segment Detection and Refinement with Deep Image Gradients},
  booktitle = {IEEE/CVF Conf. on Computer Vision and Pattern Recognition (CVPR)},
  year      = {2023}, note = {Line attraction field. arXiv:2212.07766}
}

@article{suarez2021elsed,
  author  = {Su{\'a}rez, Iago and Buenaposada, Jos{\'e} M. and Baumela, Luis},
  title   = {{ELSED}: Enhanced Line SEgment Drawing},
  journal = {Pattern Recognition},
  volume  = {127}, pages = {108619}, year = {2022},
  doi     = {10.1016/j.patcog.2022.108619}, note = {arXiv:2108.03144}
}

@inproceedings{mirzadeh2020takd,
  author    = {Mirzadeh, Seyed Iman and Farajtabar, Mehrdad and Li, Ang and Levine, Nir and Matsukawa, Akihiro and Ghasemzadeh, Hassan},
  title     = {Improved Knowledge Distillation via Teacher Assistant},
  booktitle = {AAAI Conf. on Artificial Intelligence}, year = {2020}, note = {capacity-gap effect}
}

@article{obszarski2025quant,
  author  = {Tumialis, Pawe{\l} and Skierkowski, Marcel and Przychodny, Jakub and Obszarski, Pawe{\l}},
  title   = {The Impact of 8- and 4-Bit Quantization on the Accuracy and Silicon Area Footprint of Tiny Neural Networks},
  journal = {Electronics}, volume = {14}, number = {1}, pages = {14}, year = {2025},
  note    = {DOI:10.3390/electronics14010014; closest methodological twin (8- vs 4-bit, KD+QAT).}
}

@misc{linpu_warm_kaggle,
  author       = {Hassani Shariat Panahi, Parsa},
  title        = {{NPLSD-H} warm-start training and evaluation notebook},
  howpublished = {\url{https://www.kaggle.com/code/parsahshariatpanahi/linpu-warm}},
  year         = {2026},
  note         = {Kaggle notebook; reproduces the warm-started accuracy figures of this paper}
}

@misc{mlsdn6_kaggle,
  author       = {Hassani Shariat Panahi, Parsa},
  title        = {{NPLSD-M} training notebook: an accelerator-compatible adaptation of the {M-LSD} trunk},
  howpublished = {\url{https://www.kaggle.com/code/parsahshariatpanahi/mlsd-n6-compatible}},
  year         = {2026},
  note         = {Kaggle notebook; identical head, loss, and recipe as {NPLSD-H}}
}

@misc{lsd_dst_kaggle,
  author       = {Hassani Shariat Panahi, Parsa},
  title        = {{LSD-DST}: Wireframe training data and {ImageNet}-pretrained trunk weights},
  howpublished = {\url{https://www.kaggle.com/datasets/parsahshariatpanahi/lsd-dst}},
  year         = {2026},
  note         = {Dataset; includes hgnetv2\_b0\_stage1 and mobilenetv2\_imagenet\_enc warm-start checkpoints}
}

@misc{jalilvand2026lowlatency,
	title={A Low-Latency ASIC Architecture for Real-Time Line Segment Detection}, 
	author={Amir Hossein Jalilvand and Parsa Hassani Shariat Panahi and M. Hassan Najafi},
	year={2026},
	eprint={2608.06439},
	archivePrefix={arXiv},
	primaryClass={cs.AR},
	url={https://arxiv.org/abs/2608.06439}, 
}

\end{document}